\documentclass[12pt,preprint]{aastex}

\def\pt{{p_t}}
\def\pr{{p_r}}
\def\ph{{p_\theta}}
\def\pf{{p_\phi}}

\def\Fx{{\bf x}}
\def\Fp{{\bf p}}
\def\Fv{{\bf v}}

\def\Hkep{{H_{\rm Kep}}}
\def\Hs{{H_{\rm S}}}
\def\Hlt{{H_{\rm LT}}}
\def\apo{_{\rm apo}}
\def\peri{_{\rm peri}}
\def\Vgal{{V_{\rm Gal}}}

\def\Ha{H_A}
\def\Hb{H_B}

\def\pderiv(#1/#2){\frac{\partial#1}{\partial#2}}
\def\der#1#2{\frac{d#1}{d#2}}

\def\dd{\displaystyle}

\shorttitle{}
\shortauthors{}

\begin{document}

\title{On post-Newtonian orbits \\ and the Galactic-center stars}
\author{Miguel Preto}
\affil{Astronomisches Rechen-Institut, Zentrum f\"{u}r Astronomie, \\
University of Heidelberg, D-69120 Heidelberg, Germany}
\and
\author{Prasenjit Saha}
\affil{Institute for Theoretical Physics, University of Z\"{u}rich, \\
Winterthurerstrasse 190, CH-8057 Z\"{u}rich, Switzerland}

\begin{abstract}
  Stars near the Galactic center reach a few percent of light speed
  during pericenter passage, which makes post-Newtonian effects
  potentially detectable.  We formulate the orbit equations in
  Hamiltonian form such that the $O(v^2/c^2)$ and $O(v^3/c^3)$
  post-Newtonian effects of the Kerr metric appear as a simple
  generalization of the Kepler problem. A related perturbative
  Hamiltonian applies to photon paths. We then derive a symplectic
  integrator with adaptive time-steps, for fast and accurate numerical
  calculation of post-Newtonian effects.  Using this integrator, we
  explore relativistic effects.  Taking the star S2 as an example, we
  find that general relativity would contribute tenths of mas in
  astrometry and tens of $\rm km\;s^{-1}$ in kinematics.  (For
  eventual comparison with observations, redshift and time-delay
  contributions from the gravitational field on light paths will need
  to be calculated, but we do attempt these in the present paper.) The
  contribution from stars, gas, and dark matter in the Galactic center
  region is still poorly constrained observationally, but current
  models suggest that the resulting Newtonian perturbation on the
  orbits could plausibly be of the same order as the relativistic
  effects for stars with semi-major axes $\gtrsim 0.01$~pc (or
  250~mas).  Nevertheless, the known and distinctive {\it time
  dependence\/} of the relativistic perturbations may make it possible
  to disentangle and extract both effects from observations.
\end{abstract}

\keywords{galaxy: center, relativity, stellar dynamics, methods: numerical}

\section{Introduction}
The Galactic-center stars are a population of fast-moving stars in
highly eccentric nearly Keplerian orbits around a compact mass ---
presumably a massive black hole (henceforth MBH) --- of mass
$\simeq4\times10^6M_\odot$
\citep{2005ApJ...620..744G,2008ApJ...689.1044G,2009ApJ...692.1075G}.
Pericenter distance for some of these stars are inferred to be $<10^4$
of the gravitational radius ($\simeq 3\times10^3$ in the case of S2)
implying pericenter velocities of a few percent of light.

The high pericenter velocities inspire a search for general
relativistic perturbations to Keplerian orbits.
\cite{1998AcA....48..653J} and \cite{2000ApJ...542..328F} suggested
that the precession of the pericenter, which is an effect of $O(v^2)$
where $v$ is the pericenter velocity in light units, would be
observable.  Interferometric instruments currently under development
\citep[see, for example,][]{2009svlt.conf..361E} make this possibility
more likely.  If stars even further in are discovered, then $O(v^4)$
could become detectable from astrometry, leading to tests of no-hair
theorems \citep{2008ApJ...674L..25W}.  Other consequences of general
relativity may become accessible to spectroscopic observations.
\cite{2006ApJ...639L..21Z} note that the $O(v^2)$ effect of
gravitational redshift may be within the reach of current instruments.
\cite{2009ApJ...690.1553K} suggest that the next generation of
spectrographs may pick up even the $O(v^4)$ signature of frame
dragging for the known Galactic-center stars.

To study the relativistic perturbations in detail, good methods for
computing them numerically are highly desirable.  In this paper we
provide such a method.

Since geodesics in a Kerr metric are integrable \citep[see, for
example, chapter~7 of][]{1983mtbh.book.....C}, our problem may at
first seem a trivial one.  But the known integrability only puts the
solution in terms of quadratures, it does not provide solutions in
terms of elementary functions.  The explicit forms given by
\cite{2007CQGra..24.1775K} for polar orbits, and the numerical methods
given by \cite{2009arXiv0903.0620D} for computing null geodesics give
an idea of the complexity of the formally exact solutions.  Moreover,
even if one had the exact Kerr geodesics, one would need to perturb
them for the Galactic-center stars, because of masses other than the
black hole.

In this paper we take a different route.  We go back to the Kerr
metric and derive a post-Newtonian Hamiltonian, where relativistic
effects appear as perturbations to Keplerian orbits and to which
Newtonian perturbations due to other masses can be trivially added.
Then we consider numerical algorithms designed for comets and other
highly eccentric orbits in the solar system, and generalize them to
work for post-Newtonian perturbations as well.

\section{A post-Newtonian Hamiltonian}

Geodesic equations in a given metric can be described in terms of a
Lagrangian
\begin{equation}
L = {\textstyle\frac12} \, g_{\mu\nu} \, \der{x^\mu}\tau \der{x^\nu}\tau,
\end{equation}
where $\tau$ is the affine parameter which, conveniently for us, can
be identified with the star's proper time.  Less common is the 
equivalent Hamiltonian form
\begin{equation}
H = {\textstyle\frac12} \, g^{\mu\nu} \, p_\mu p_\nu.
\end{equation}
In fact $H$ and $L$ are equal, and conserved along geodesics.  The above
duality is simply a consequence of being quadratic.

We want to derive a post-Newtonian Hamiltonian for a test particle in
a Kerr metric.  To do this, we write the metric in Boyer-Lindquist
coordinates, adopting $(-,+,+,+)$ for the signature of the metric and
choosing units in which $G=c=1$, as follows:
\begin{equation}
g_{\mu\nu}=
\left(
\begin{array}{cccc}
\dd -\left(1-\frac{2\mu}\Sigma\right) & ~ & ~
 & \dd -\frac{2\mu\sigma}\Sigma r\sin\theta \\
~ & \qquad \dd \frac\Sigma\Delta \qquad & ~ & ~ \\
~ & ~ & \qquad \dd \Sigma r^2 \qquad & ~ \\
\dd -\frac{2\mu\sigma}\Sigma r\sin\theta & ~ & ~
& \dd \left(\Delta+\frac{2\mu}\Sigma\Big(\Delta+2\mu\Big)\right)
  r^2\sin^2\theta \\
\end{array}
\right).
\end{equation}
Here we have defined
\begin{equation}
\mu = \frac Mr \qquad \sigma = \frac sr \sin\theta \qquad
\kappa = \frac sr \cos\theta
\end{equation}
and
\begin{equation}
\Sigma \equiv 1 + \kappa^2 \qquad
\Delta \equiv 1 - 2\mu + \kappa^2 + \sigma^2,
\end{equation}
where the dimensionless spin parameter $s\in[0,1]$. 
The contravariant components are, as is not too difficult to 
verify by inverting the matrix, as follows:
\begin{equation}
g^{\mu\nu}=
\left(
\begin{array}{cccc}
\dd - \left(1+\frac{2\mu}{\Sigma\Delta}\Big(\Delta+2\mu\Big)\right)
 & ~ & ~
 & \dd -\frac{2\mu\sigma}{\Sigma\Delta\, r\sin\theta} \\
~ & \qquad \dd \frac\Delta\Sigma \qquad & ~ & ~ \\
~ & ~ & \qquad \dd \frac1{\Sigma r^2} \qquad & ~ \\
\dd -\frac{2\mu\sigma}{\Sigma\Delta\, r\sin\theta} & ~ & ~
& \dd \frac{1-\sigma^2/\Delta}{\Sigma\Delta\, r^2\sin^2\theta} \\
\end{array}
\right).
\end{equation}
Expanding to second order in $\mu,$ $\sigma,$ and $\kappa$ we have
\begin{equation}
g^{\mu\nu}=
\left(
\begin{array}{cccc}
-(1+2\mu+4\mu^2) & ~ & ~  & \dd -\frac{2\mu\sigma}{r\sin\theta} \\
~ & \qquad 1-2\mu+\sigma^2 \qquad & ~ & ~ \\
~ & ~ & \dd \qquad \frac{1-\kappa^2}{r^2} \qquad & ~ \\
\dd -\frac{2\mu\sigma}{r\sin\theta} & ~ & ~ & \dd \frac1{r^2\sin^2\theta} \\
\end{array}
\right).
\end{equation}

Now we proceed to consider the Hamiltonian.  For simplicity, we put
$M=1$, which just means we are measuring $r$ in units of the gravitational 
radius, $GM/c^2$.  To consider dynamics at large-$r$ we replace
\footnote{Here $\epsilon$ is just a label for keeping track of
  orders.  Numerically $\epsilon=1$.}
\begin{equation}
   r \rightarrow \epsilon^{-2} r.
\end{equation}
The Hamiltonian will now have two regimes.

In the low-velocity regime we require the velocity terms to be
$O(\epsilon)$.  This is achieved with the substitutions
\begin{equation}
   \pr \rightarrow \epsilon \pr        \qquad
   \ph \rightarrow \epsilon^{-1} \ph   \qquad
   \pf \rightarrow \epsilon^{-1} \pf
\end{equation}
which gives
\begin{equation}
H = - \frac{\pt^2}2 + \left( \frac{\pr^2}2 + \frac{\ph^2}{2r^2} +
      \frac{\pf^2}{2r^2\sin^2\theta} - \frac{\pt^2}r \right) \epsilon^2
    - \left( \frac{2\pt^2}{r^2} + \frac{\pr^2}r \right) \epsilon^4 
    - \frac{2s\pt\pf}{r^3} \epsilon^5 
    + O(\epsilon^6).
\label{Hpn}
\end{equation}
At zeroth order, a test particle just stays still. At $O(\epsilon^2)$
it follows Newtonian dynamics.  Relativistic effects appear at
$O(\epsilon^4)$, while frame-dragging appears at $O(\epsilon^5)$. 
But note that the kinematic effects themselves are at one order of
$\epsilon$ lower: thus Newtonian velocities are $O(\epsilon)$,
Schwarzschild perturbations to the velocities are $O(\epsilon^3)$,
while the frame-dragging perturbation to velocity is $O(\epsilon^4)$.
Gravitational radiation is a higher-order effect which we disregard
here, and in any case the timescale for orbital decay of the S2 star
due to radiation reaction is longer than a Hubble time.

In the light-velocity regime we require the velocity terms to be
$O(1)$. Thus we replace
\begin{equation}
   \ph \rightarrow \epsilon^{-2} \ph   \qquad
   \pf \rightarrow \epsilon^{-2} \pf
\end{equation}
which gives
\begin{eqnarray}
H &=& - \frac{\pt^2}2 + \frac{\pr^2}2 + \frac{\ph^2}{2r^2}
    + \frac{\pf^2}{2r^2\sin^2\theta}
    - \left( \frac{\pt^2}r + \frac{\pr^2}r \right) \epsilon^2 + \nonumber \\
  & & - \left( \frac{2\pt^2}{r^2}
             - \frac{s^2\sin^2\theta}{2r^2} p_r^2
             + \frac{s^2\cos^2\theta}{2r^4} p_\theta^2
             + \frac{2s\pt\pf}{r^3}
      \right) \epsilon^4 + O(\epsilon^6).
\label{Hpn-null}
\end{eqnarray}
At zeroth order, null geodesics just move in straight lines.  The
leading perturbation is at $O(\epsilon^2)$.

The Hamiltonian (\ref{Hpn}) is the approximate Hamiltonian we will
use, but we can simplify its form with some variable changes.  First,
we set $\pt=-1$, which we are free to do since the Hamiltonian is 
autonomous.  This merely sets units for the affine parameter
such that $dt/d\tau=-1$ in the large-$r$ limit, and has no physical
significance.  Then, we change from $r,\theta,\phi$ to $x,y,z$.
Completing the canonical transformation, we have
\begin{equation}
\pr = \frac{\Fx\cdot\Fp}{r}, \qquad
\pf = (\Fx\times\Fp)_z,
\end{equation}
and hence
\begin{equation}
\pr^2 + \frac{\ph^2}{r^2} + \frac{\pf^2}{r^2\sin^2\theta} = \Fp^2.
\end{equation}
The post-Newtonian Hamiltonian (\ref{Hpn}) then becomes
\begin{equation}
H = \Hkep + \Hs + \Hlt,
\label{Hpn1}
\end{equation}
where
\begin{eqnarray}
\label{Hpnf}
\Hkep &=& \frac{\Fp^2}2 - \frac1r, \nonumber \\
\Hs   &=& - \frac2{r^2} - \frac{(\Fx\cdot\Fp)^2}{r^3}, \\
\Hlt  &=& 2\, \frac{{\bf s}\cdot\Fx\times\Fp}{r^3}. \nonumber
\end{eqnarray}
There is a separate equation for $t$
\begin{equation}
\dot t = 1 + \frac2r + \frac4{r^2}  - \frac{2s\pf}{r^3}.
\end{equation}

\section{An adaptive-timestep symplectic integrator}

The post-Newtonian orbit equations can be integrated numerically by
any general-purpose method for ordinary differential equations.
Another option is to use an $N$-body simulation code for dense stellar
systems, with post-Newtonian terms added \citep{2008AJ....135.2398M}.
A computationally more efficient strategy, however, would be an
integration algorithm that takes advantage of the above formulation of
relativistic effects as small perturbation to a Kepler Hamiltonian.
We now design such an integration algorithm, based on recent work on
cometary orbits, which are also highly eccentric orbits that
experience interesting perturbations around pericenter passage.

When integrating Hamiltonian systems numerically, it is a common
practice to impose the condition that the numerical solution has (to
machine precision) the symmetry properties of Hamiltonian flow.
Integration algorithms with this property are known as symplectic
integrators.  A simple but important example is generalized
leapfrog. Suppose we have a Hamiltonian that is the sum of two parts
$H=\Ha+\Hb$ where $\Ha$ and $\Hb$ are individually easy or trivial to
integrate. Generalized leapfrog evolves under $H$ for a time step
$\Delta\tau$ as follows.
\begin{enumerate}
\item Evolve under $\Ha$ for time $\frac12\Delta\tau$.
\item Evolve under $\Hb$ for time $\Delta\tau$.
\item Reiterate step 1.
\end{enumerate}
If $H=\frac12\Fp^2+V(\Fx)$ the above becomes the classical leapfrog
integrator.  It turns out \citep[see, for example,][]{1990Forest} that
generalized leapfrog amounts to evolving under a ``surrogate''
Hamiltonian
\begin{equation}
   \Ha + \Hb + \Big(
   {\textstyle\frac1{12}} \{\Ha,\Hb\},\Hb\} +
   {\textstyle\frac1{24}} \{\Ha,\Hb\},\Ha\}
   \Big) \Delta\tau^2 + O(\Delta\tau^4) .
\label{Hsurr}
\end{equation}
The nested Poisson brackets amount to a Hamiltonian expression for the
error, which is manifestly second order.  Higher-order extensions are
possible \citep{1990Forest,1990Yoshida,2001CeMDA..80...39L} but in
practice second-order is the most used.  If one of $\Ha$ or $\Hb$ is
much smaller than the other, the error Hamiltonian will be
correspondingly small.  \cite{1991AJ....102.1528W} and independently
\cite{1991CeMDA..50...59K} proposed integrators for planetary orbits
where $\Ha$ is the integrable Kepler Hamiltonian and $\Hb$
encapsulates the perturbations.  For planetary orbits, the low order
of generalized leapfrog becomes an advantage in that the steps can be
made comparatively large ($\approx10$ steps per orbit) and still
provide high accuracy.  Further refinements are possible, such as
perturbative pre-processing of the initial conditions
\citep{1992AJ....104.1633S} or perturbative post-processing of the
results \citep{1996FIC....10..217W}, but the original Wisdom-Holman
scheme is the most common choice for long-term solar-system orbit
integrations.

Another kind of symplectic integrator, also second order, is the
implicit midpoint method, due to \cite{1986Feng}.  This can be written
as a simple discretization of Hamilton's equations
\begin{equation}
\Delta \Fx =   \left( \pderiv(H/\Fp) \right) \Delta \tau  \qquad
\Delta \Fp = - \left( \pderiv(H/\Fx) \right) \Delta \tau
\label{feng1}
\end{equation}
with the derivatives are evaluated at the midpoint
\begin{equation}
\left( \Fp+{\textstyle\frac12}\Delta\Fp,
       \Fx+{\textstyle\frac12}\Delta\Fx \right) .
\label{feng2}
\end{equation}
It is essential for $\Delta\Fx,\Delta\Fp$ to be consistent between
(\ref{feng1}) and (\ref{feng2}) to high accuracy (preferably machine
precision), otherwise the symplectic property is lost.  Hence, the
implicit midpoint method requires iteration.  But it requires no
splitting of the Hamiltonian, and hence is very useful when
generalized leapfrog is inapplicable.  A surrogate Hamiltonian for the
implicit midpoint integrator is derived in \cite{1997AJ....114..409S}.

Combining the two above ingredients, a possible integration method for
the Galactic-center stars would be a generalized leapfrog with
$\Ha=\Hkep$ and $\Hb=\Hs+\Hlt+\Vgal$, with an exact Kepler solution
used for the former, and implicit midpoint used for the latter.  In
fact, neither of the separate integrations under $\Ha$ and $\Hb$ needs
to be exact.  As long as they are symplectic and second-order, the
surrogate Hamiltonian (\ref{Hsurr}) will apply.

For low eccentricites, an integrator as above would be very efficient.
When we consider Galactic-center stars (or comets) however, we run
into the major limitation of generalized leapfrog: the stepsize
$\Delta\tau$ must remain fixed, otherwise the integrator is no longer
symplectic. Yet for $e\approx0.9$, the very small $\Delta\tau$ needed
at pericenter becomes hopelessly expensive if used throughout an
orbit.  Adaptive time-stepping is needed.

The key to adaptive time-stepping is to transform from $\tau$ to a new
independent variable (say $s$) which somehow implements the desired
stretching and shrinking of the stepsize without breaking the
symplectic property.  \cite{1997CeMDA..67..145M} provided the first
example, showing how a simple modification of the Wisdom-Holman
algorithm effective creates the variable $s$ with $d\tau=r\,ds$.  Then
\cite{1999AJ....118.2532P} and independently
\cite{1999CeMDA..74..287M} formulated, for a Hamiltonian
$\frac12\Fp^2+V(\Fx),$ a time transformation with $d\tau=-ds/V(\Fx)$.
As a by-product, this work produced a leapfrog for $\Hkep$ that is
symplectic and in fact recovers the exact answer except for a very
small phase error, but is computationally much simpler than the exact
solution.  \cite{2002CeMDA..84..343M} then generalized these ideas to
make the time-transformation completely adaptive, though not in a
Hamiltonian formulation, so it was not manifestly symplectic.  Later,
\cite{2007CeMDA..98..191E} supplied an elegant Hamiltonian derivation
of adaptive stepsize.

Based on all the above, we now develop our integrations algorithm.
The derivation basically follows \cite{2007CeMDA..98..191E} but is
written with a view to application to post-Newtonian orbits.

Let us enhance the phase space $(\Fx,\Fp)$. We now treat $\tau$ as an
additional coordinate, with a new variable $\Phi$ being its conjugate
momentum, and the new independent variable $s$.  In this enhanced
phase space, consider the Hamiltonian
\begin{equation}
     F(\Fx,\tau,\Fp,\Phi) =
     \frac{H-k}\Phi + \ln\left(\frac\Phi{\varphi(\tau)}\right),
\end{equation}
where $\varphi(\tau)$ is a known function and $k$ is a constant. 
Hamilton's equations for $F$ are
\begin{equation}
    \der \Fx s =  {1\over \Phi} \pderiv(H/\Fp) \qquad
    \der \Fp s = -{1\over \Phi} \pderiv(H/\Fx),
\end{equation}
together with
\begin{equation}
     \der \Phi s = \der {}\tau \ln\varphi
\label{derPhi}
\end{equation}
and
\begin{equation}
     \der \tau s = -\frac{H-k}{\Phi^2} + \frac1\Phi.
\end{equation}
Moreover, $F$ will be conserved.

Suppose at some $s$ we have $H=k$.  Combining the two previous
equations, we infer that at this point
\begin{equation}
     \der {}\tau \ln\Phi = \der {}\tau \ln\varphi.
\end{equation}
Thus, in the neighborhood of this point, $\Phi/\varphi$ will be
constant, and hence $H-k$ will remain zero.  In other words, if
\begin{equation}
     \der \tau s = \frac1\Phi, 
\end{equation}
holds initially, it will continue to hold.
The interpretation is the original Hamiltonian equations with a
rescaled variables $ds=\Phi\,d\tau$.

Now the cunning part: we consider $\varphi(\tau)$ as
$\varphi(\Fx,\Fp)$ where $\Fx,\Fp$ are functions of $\tau$, and
replace $d\varphi/d\tau$ by the convective derivative.  We have:
\begin{equation}
    \der\Phi s = \left( \der \Fx\tau \cdot \pderiv(/\Fx)
                      + \der \Fp\tau \cdot \pderiv(/\Fp) \right)
                 \ln\varphi(\Fx,\Fp)
\label{dvarphi}
\end{equation}

With the above ingredients in hand, we proceed to write a leapfrog for
$F$ in the variable $s$:
\begin{enumerate}
\item Holding $\Fx,\Fp$ constant, advance $\Phi$ using (\ref{dvarphi})
  over $\Delta s=\frac12$.  This amounts to evolution under a
  Hamiltonian $-\ln\varphi$.
\item Holding $\Phi$ constant, evolve $\Fx,\Fp$ under $H/\Phi$ for
  $\Delta s=1$, and advance $\tau$ by 1. This amounts to evolution
  under a Hamiltonian $(H-k)/\Phi+\ln\Phi$.
\item Reiterate step 1.
\end{enumerate}
In the present work, step 2 above consists of a generalized leapfrog
with the main Hamiltonian split into Keplerian, post-Newtonian, and
external potential parts. We also have two minor simplifications:
(i)~we assume $\varphi$ depends only on $\Fx$, and (ii)~we disregard
the evolution of $\tau$.  The simplified treatment of time is fine for
our test particle integrations, but for an $N$-body formulation, the
time equation needs to be treated with special care, and the form
$d\tau=ds/\Phi$, associated with the new pair of conjugate variables
($\tau,\Phi$) presents several advantages.

The algorithm is as follows.
\begin{enumerate}
\item Advance $\Phi$ by $\frac12(\partial H / \partial \Fp) \cdot
  \nabla (\ln\varphi)$.
\item Evolve $\Fx,\Fp$ under $\Hkep$ for $\Delta\tau=1/(2\Phi)$, using the
  algorithmic regularization scheme of \cite{1999AJ....118.2532P}.
\item Evolve $\Fx,\Fp$ under $\Hs+\Hlt+\Vgal$ for $\Delta\tau=1/\Phi$,
  using the implicit midpoint method.  For the small post-Newtonian
  contributions of interest for this paper, this implicit scheme
  converges to machine precision in two or there iterations.
\item Reiterate step 2.
\item Reiterate step 1.
\end{enumerate}

The following comments about the integrator are worth making.  First,
with the simplest choice $\varphi \propto 1/r$, we obtain an adaptive,
symplectic integrator to integrate eccentric weakly-perturbed
Keplerian orbits --- including non-separable post-Newtonian
perturbations.  This integrator is easy to implement, and it is free
from the instability found \cite{1999AJ....117.1087R} in
fixed-stepsize integration of highly eccentric orbits.  Second, with
the apparently innocuous modification of the time transformation to
$d\tau=ds/\Phi$, the Keplerian part becomes trivial to integrate,
since $\Phi$ is kept constant while the (cartesian) coordinates are
advanced.  This is very advantageous for a $N$-body implementation
with individual time steps, where particle synchronization requires
numerous Kepler drifts.  Third, being a symplectic integrator, errors
in the longitudes grow only linearly with time and, for spherical
perturbations, angular momentum is conserved to machine
precision. This is clearly a most desirable property --- in
particular, in dynamical problems for which long term integrations
with accurate tracking of all phase angles are required, {\em e.g.,}
resonant relaxation and related effects
\citep{2007MNRAS.379.1083G,2009ApJ...697L..44M,perets09}.  Fourth, as pointed
out already by \cite{2007CeMDA..98..191E}, the freedom to choose the
functional form of $\varphi$ provides an additional degree of freedom
that can be explored in order to resolve close encounters without
breaking symplecticity.  Fifth, this integrator and the post-Newtonian
approximation presented in this work are very well suited to be
included in orbital fitting routines of the Galactic-center stars
\citep{2005ApJ...622..878W,2009ApJ...692.1075G}. Sixth, this symplectic
integrator is also ideally suited for implementation in view of $N$-body 
modelling of gravitational wave sources, {\it e.g.} extreme mass ratio 
inspirals, which are of great interest  for LISA (Laser Interferometer Space 
Antenna). 

\section{Numerical results}

As an illustration, we consider perturbations of the orbit
with Keplerian elements
\begin{equation}
a = 2.4\times10^4 \qquad e = 0.88 \qquad
I = 135.25^\circ \qquad \Omega = -134.71^\circ \qquad \omega = 63.56^\circ
\label{S2-like}
\end{equation}
which are approximately the measured values for S2
\citep{2009ApJ...692.1075G}.  Note that $a$ is in units of the
gravitational radius $GM/c^2\simeq5\times10^6\,\rm km$ or
$\simeq4\mu\rm as$ on the sky.  We assume the spin is unity (maximal)
and directed along $+z$.

\begin{figure}
\epsscale{0.8}
\plotone{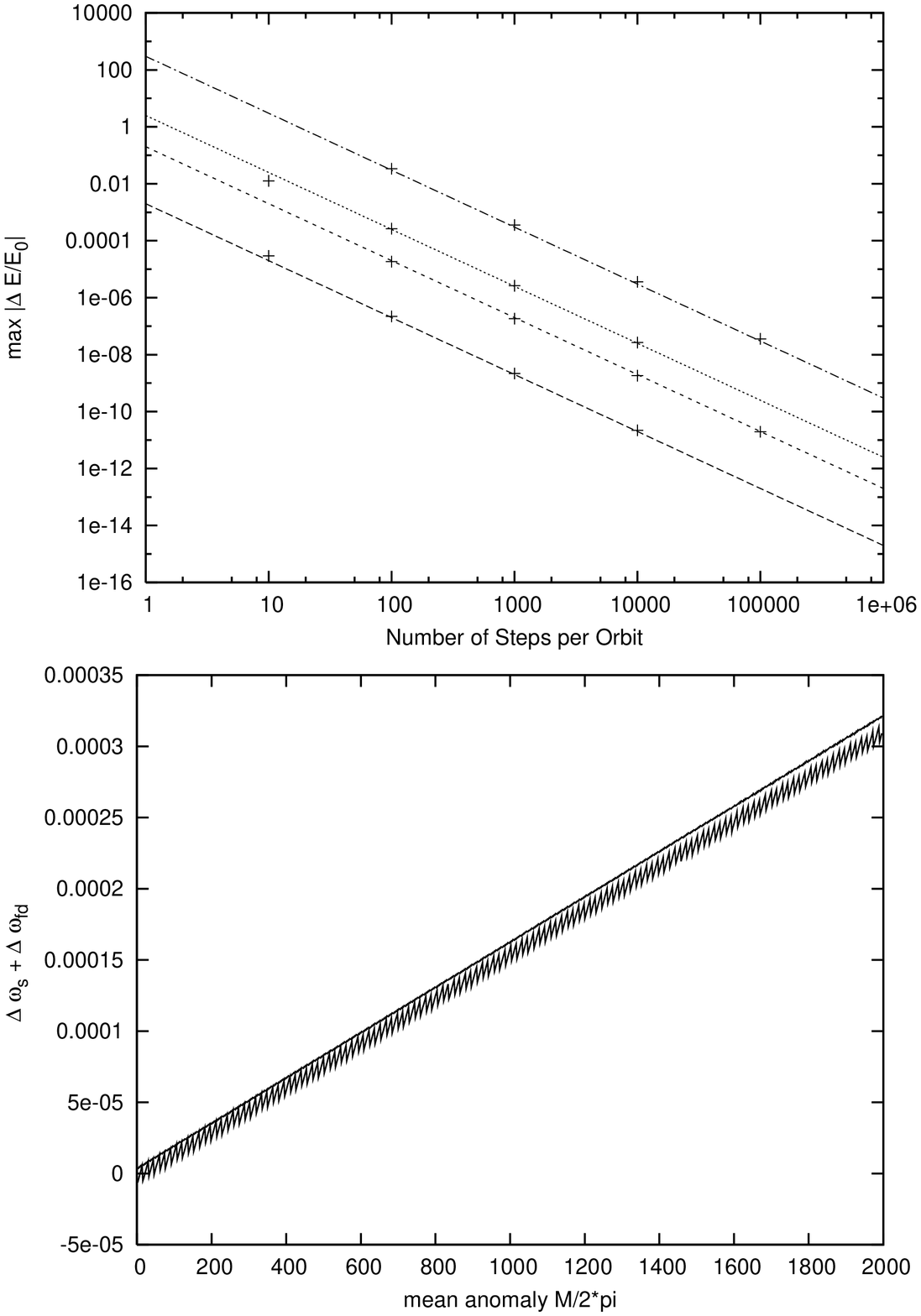}
\caption{Verification of the properties of the integrator.  The upper
panel shows the maximum error in $H$ over 2000 periods for four orbits
with $a=2.4\times10^4$ and $e=0.5, 0.88, 0.95$ and $0.99$. The lower
panel shows the pericenter angle difference between the integration of an 
S2-like orbit,  with initial time steps $1/\Phi(0)$ being $10^{-3}$ and $10^{-4}$ 
of the Keplerian period, and the theoretical value in~(\ref{weinb}). The oscillations were 
averaged out from the time series to highlight its (linear) secular evolution, 
although a small oscillation remains in the former case.}
\label{fig-errors}
\end{figure}

We begin by verifying that the integrator is indeed second order and
that the phase error grows linearly with time.  From
Figure~\ref{fig-errors} we see that the maximum error in $H$ is
proportional to $\Phi^{-2}(0)$, where $1/\Phi(0)$ is the initial
stepsize, and that the error in pericenter angle is proportional to the 
number of orbits ({\it i.e.} it is linear rather quadratic). In order to evaluate 
this error we adopt the expressions for the shift of pericenter due to a Schwarzschild
black hole up to $2^{nd}$ order \citep{1972gcpa.book.....W,2007PASP..119..349N} 
and the $1^{st}$ order contribution from the spin \citep{2009ApJ...690.1553K}:
\begin{eqnarray}
\Delta \omega_s & = & \frac{6 \pi}{a(1-e^2)} + \frac{3 \pi (18+e^2)}{2 a^2 (1-e^2)^2} \nonumber \\
                                 &  &  \label{weinb}   \\
\Delta \omega_{fd} & = & -\frac{12 \pi s \cos I}{a^{3/2} (1-e^2)^{3/2}}.   
\nonumber
\end{eqnarray}

The observable orbit, that is to say, the sky position and redshift as
a function of observer time, depends also on the light path from the
star to the observer.  There are two types of effects.
\begin{enumerate}
\item The sky position of the star will be slightly shifted by
  gravitational lensing.  The maximum lensing displacement is the
  Einstein radius $R_E$.  Since for Galactic-Center stars, the
  lens-source distance $D_{LS}$ is much smaller than the
  observer-source and lens-source distances, the usual expression for
  $R_E$ simplifies to
\begin{equation}
R_E = 2\sqrt{D_{LS}}
\end{equation}
in units of the gravitational radius, and $D_{LS}$ will be of order $a$.
\item Then there is the redshift, and the R\o mer effect, which can
  be considered as the redshift integrated along the orbit.  The
  $O(\epsilon)$ contribution to the redshift is classical. Special
  relativity and the time part of the metric both contribute to the
  redshift at $O(\epsilon^2)$, while space curvature contributes at
  $O(\epsilon^3)$.
\end{enumerate}
We do not include all these effects in this paper, leaving it for
future work, because at this stage our aim is to gain some insight
into the size of relativistic effects rather than calculate
observables precisely. However, following \cite{2006ApJ...639L..21Z}, 
we include the $O(v^2)$ effects due to gravitational time dilation
when we estimate the perturbations on the radial velocities.

\begin{figure}
\epsscale{0.8}
\plotone{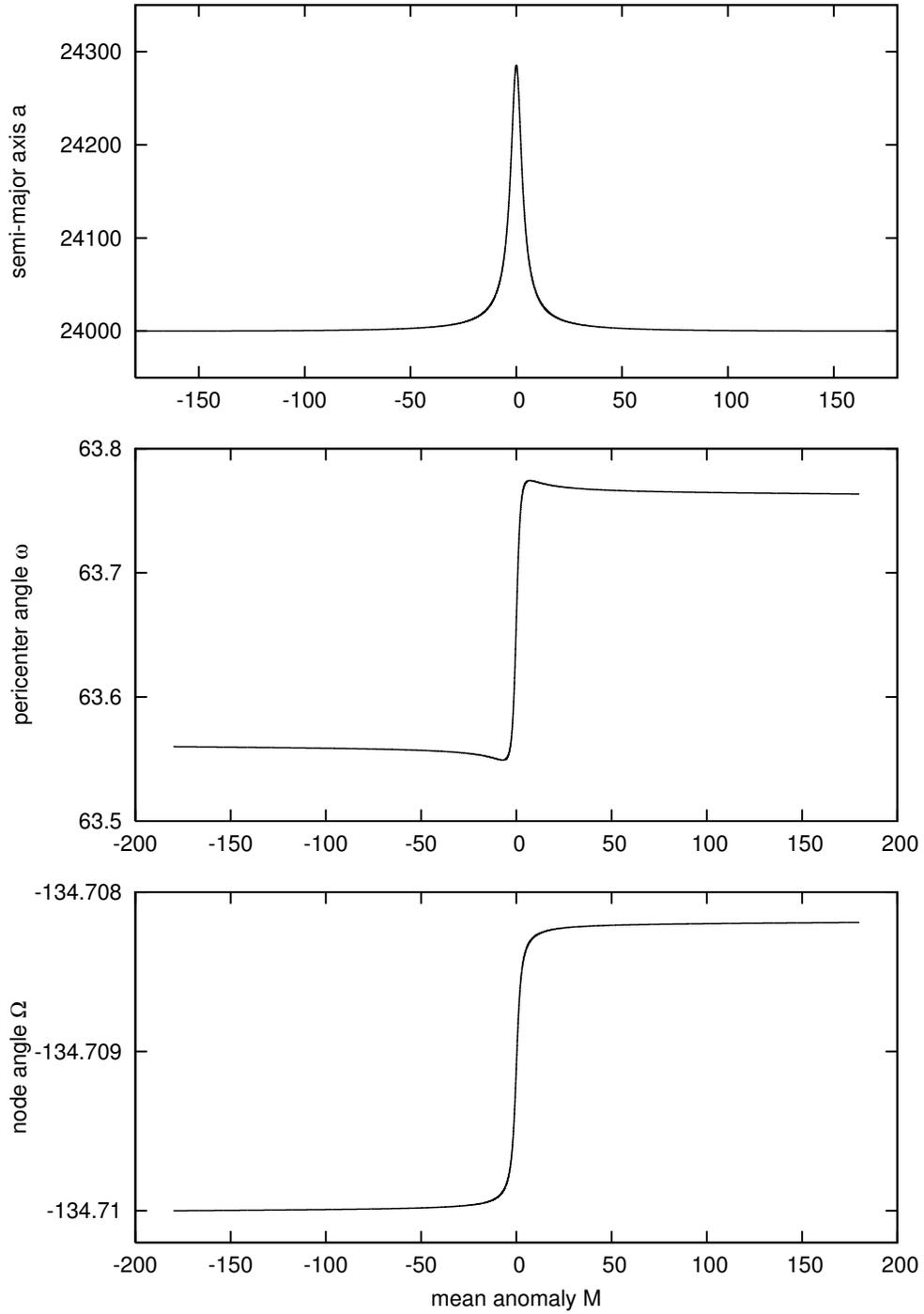}
\caption{Perturbation of the (osculating) orbital elements of an
  S2-like orbit (see Eq.~\ref{S2-like}) due to post-Newtonian terms.
  The upper panel shows $a$ in gravitational units, the middle panel
  and lower panels show $\omega$ and $\Omega$.  The mean anomaly is
  taken as the independent variable, as a surrogate for time (see
  text). All angles are in degrees.}
\label{fig-elem}
\end{figure}

Having chosen the orbital elements, we start the star at apocenter and
integrate the orbit under the Hamiltonian (\ref{Hpn1}) to the next
apocenter passage.  Figure~\ref{fig-elem} shows $a$, $\omega$, and
$\Omega$ along the orbit, as a function of the mean anomaly, which is
a surrogate for time.  The range $-180$ to 180 in the mean anomaly
corresponds to a Keplerian orbit going from apocenter to apocenter.
In this paper we always compare perturbed and unperturbed orbits for
the same value of mean anomaly, not necessarily the same value of
time.  If time is taken as the independent variable, an artificial
secular drift would appear, because the perturbed orbit has a slightly
different orbital period.  The other two Keplerian elements are not
shown here, because $I$ and $a(1-e^2)$ are constant as a consequence
of the conservation of total angular momentum, which the symplectic
integrator reproduces to roundoff error.

From Figure~\ref{fig-elem}, one easily verifies the well-known
leading-order expressions
\begin{equation}
\Delta\omega = {6\pi\over a(1-e^2)}
\end{equation}
for the pericenter precession and
\begin{equation}
\Delta\Omega = -{8\pi\cos I \over \left(a(1-e^2)\right)^{3/2} }
\end{equation}
for Lense-Thirring node precession.  The maximum change in $a$ can be
estimated as follows.  We start by noting from the inspection of  (\ref{Hpnf}) 
that the leading post-Newtonian perturbation is
\begin{equation}
\Delta H = \frac2{r^2}
\end{equation}
at both pericenter and apocenter. Since at these points $r=a(1\pm e)$ we
have
\begin{equation}
\Delta H\peri - \Delta H\apo = \frac{8e}{a^2(1-e^2)^2}
\end{equation}
For the unperturbed Hamiltonian $H=-1/(2a)$ hence $\Delta H\simeq
\Delta a/(2a^2)$ from which it follows
\begin{equation}
\Delta a = \frac{16e}{(1-e^2)^2}
\end{equation}

Using similar arguments, or dimensional analysis, one easily derives
that the effects of a perturbation
\begin{equation}
\Delta H \sim r^{-n}
\label{potscaling}
\end{equation}
scale as follows.
\begin{equation}
\begin{array}{lll}
\mbox{astrometric} & \Delta l,\, \Delta b       & \sim a^{2-n} \\
\mbox{kinematic}   & \Delta v                   & \sim a^{\frac12-n} \\
\mbox{precession}  & \Delta\omega,\, \mbox{etc} & \sim a^{1-n}
\end{array}
\label{effectscaling}
\end{equation}
Post-Newtonian effects have $n=2$ (Schwarzschild) or $n=3$
(Lense-Thirring).  Thus, the astrometric effect of Schwarzschild
perturbations is independent of $a$ to leading order, while other
post-Newtonian effects get stronger as orbits get smaller.
Relativistic prograde precession would seem easier to measure at
larger distances, but that is not the case since the orbital period
increases as $a^{3/2}$. On the other hand, Galactic perturbations from
other masses than the black hole would have $n<1$, and hence get
weaker as orbits get smaller.

The form and strength of Newtonian Galactic perturbations, due to
other stars, gas, and dark matter, are still poorly constrained at present.
Observations aimed at measuring post-Newtonian effects would need to
fit the local Galactic potential as well. Stars, however, are expected to be the
dominant component of the extended galactic mass distribution near the
central MBH. From observations, the stellar distribution
around the MBH is best approximated by a double power-law density profile,
with a (somewhat uncertain) break radius that may range from $r_b\sim0.1-0.2$ pc
up to $r\sim$ few$\times 1$pc \citep{2007AAP...469.125S,2009arXiv0902.3892S}. 
This translates into a single power-law model $\rho(r)=\rho_0(r/r_0)^{-\gamma}$
throughout the whole region of special interest for us, $r\lesssim0.01$pc.
Therefore, we adopt the following gravitational potential $\Vgal$

\begin{eqnarray}
\Vgal = \left\{\begin{array}{ll}
  \displaystyle{  \frac{4\pi}{(3-\gamma)(2-\gamma)} \rho_0 r_0^\gamma r^{2-\gamma} = \frac{M_*(r_0)}{(2-\gamma) r_0} \left(\frac{r}{r_0}\right)^{2-\gamma}  }, &  \gamma \ne 2,  \\ 
  & \\
  \displaystyle{ 4\pi \rho_0 r_0^2 \ln\left(\frac{r}{r_0}\right) = \frac{(3-\gamma)M_*(r_0)}{r_0} \ln\left(\frac{r}{r_0}\right) },  & \gamma=2, 
\end{array}
\right.
\label{newtonp}
\end{eqnarray}
where $M_*(r_0)$ is the total stellar mass within the radius $r_0$. We will adopt
$r_0=0.01$pc in this paper.\footnote{It is easy to see that $n=\gamma-2$, so
that $1/2 \leq \gamma < 3$ implies $-3/2 \leq n <1/2$ in (\ref{potscaling}) and
(\ref{effectscaling}).}

In the absence of definite observational measurements, and in order to
chose the values for the slope $\gamma$ and for the normalization of
the mass distribution, we have to appeal to stellar dynamical
theory. The relaxation time in the Milky Way's nucleus is $T_R\sim
O(1$Gyr) and therefore old stellar populations may have had enough
time to reach a relaxed steady-state. Mass segregation around a MBH
leads to steeper profiles for heavy stars ({\it e.g.} compact remnants
such stellar black holes) than for light stars
\citep{1977ApJ...216..883B}. Recent Fokker-Planck and $N$-body studies
\citep{2009ApJ...697.1861A, preto_amaro09} show that mass segregation is
indeed a generic and robust property of the relaxed populations around
a MBH; furthermore, it is stronger than expected according to Bahcall
\& Wolf, and leads to power-law density profiles with slopes such as
$\gamma_{{\rm heavy}}\sim1.8-2.3$ and $\gamma_{{\rm
light}}\sim1.0-1.6$. According to the latter studies, the total amount
of stellar mass packed inside $r_0=0.01$pc is $M_*(r_0) \sim \alpha
\times 2 \times 10^3 M_\odot$, where $\alpha=O(1)$. Therefore, in our
numerical tests, we will adopt $\gamma=1.5, 2.1$ and a normalization
for the total stellar mass $M_*(r_0)=2 \times 10^{3-4} M_\odot$. This
leads to circular velocities, induced by stellar mass alone, of order
20--50~km/s. Current observational constraints on the mass
normalization are still roughly one or two orders of magnitude above
of these values, whereas $\gamma$ is still very weakly constrained
\citep{2008ApJ...689.1044G, 2009ApJ...692.1075G}.

\begin{figure}
\epsscale{0.8}
\plotone{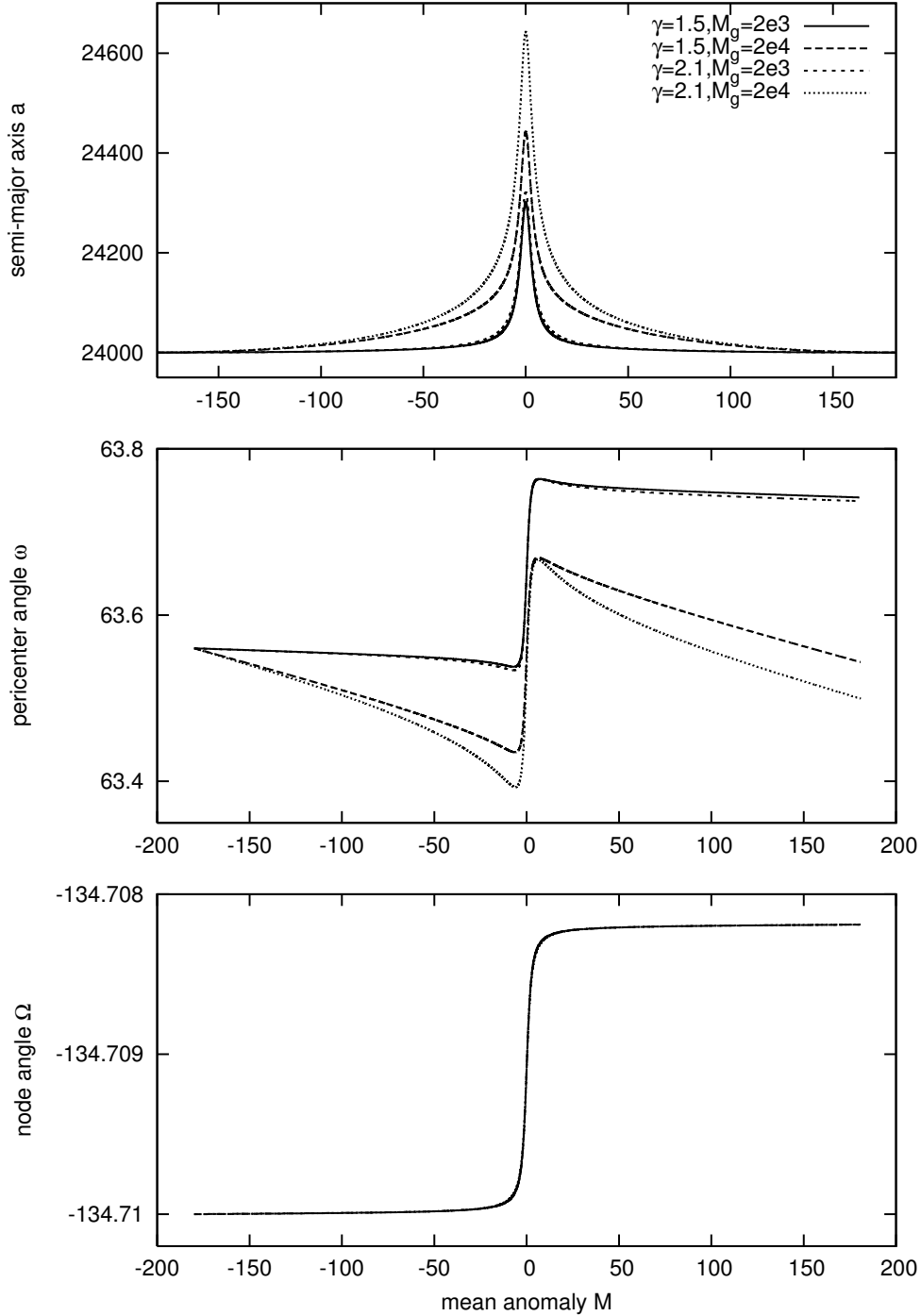}
\caption{Perturbation of the orbital elements of a S2-like orbit due
  to post-Newtonian terms plus a model for Galactic perturbations
  consisting of a stellar cluster around Sgr~A$^*$ (see
  Eq.~\ref{newtonp}). The panels are analogous to the two upper panels
  in Figure~\ref{fig-elem}.  For the Galactic contribution, the larger
  the stellar-cluster and the steeper the cusp, the stronger the
  perturbation of the elements. The assumed total mass for the
  stellar cluster is, according to current theoretical estimates,
  relatively high.}
\label{fig-elem-gal}
\end{figure}

\begin{figure}
\epsscale{0.8}
\plotone{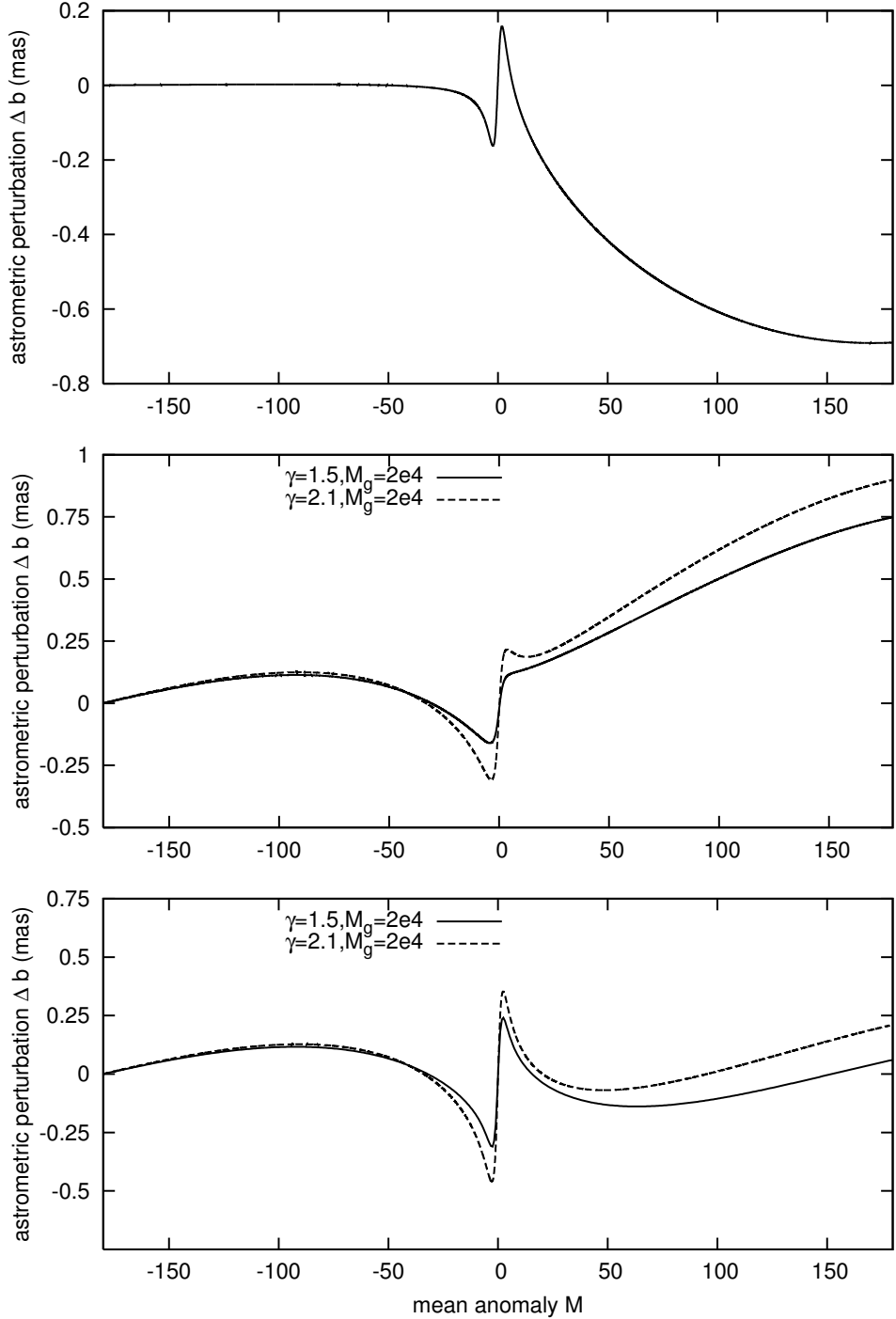}
\caption{Astrometric perturbations for an S2-like orbit.  The upper
  panel shows $\Delta b$ from post-Newtonian effects only, the middle
  panel from the model Galactic perturbations only, while the lower
  panel combines both perturbations.  The precession due to both the 
  extended mass distribution and the relativistic effects is too small to 
  be detected with current astrometric capabilities, even though the 
  stellar cusp is relatively massive.}
\label{fig-astrom}
\end{figure}

Figures \ref{fig-elem-gal} and \ref{fig-astrom} illustrate astrometric perturbations 
from the relativistic terms and from the local Galactic model (\ref{newtonp}). Not 
surprisingly, for an orbit with semi-major axis $a$ the galaxy's perturbation effects 
are $\propto M_*(a)$. For the adopted parameters, the cusp slope has only a weak 
effect but it is noticeable; furthermore, the galaxy's pertubations are also stronger for 
steeper cusps.  Given the model adopted for the stellar cluster, the cumulative mass
distribution is $M(r) = M_*(r_0) (r/r_0)^{3-\gamma}$, so when the slope $\gamma$ changes by 
an amount $\Delta \gamma$, the total mass within $a$ (orbital semi-major axis) changes by 
\begin{equation}
\Delta M(a) = M_*(r_0) \left[ \left(\frac{a}{r_0}\right)^{3-\gamma-\Delta \gamma} -  \left(\frac{a}{r_0}\right)^{3-\gamma} \right]. 
\end{equation}
Therefore, increasing $\gamma=1.5$ to $2.1$ leads to a mass increment
within $a$, $\Delta M_*(<a) \sim 0.19 M_*(a) \sim 3.8 \times 10^3
M_\odot$. Therefore a larger proper motion obtains which is, to first
order, given by $\Delta l_{2.1} \sim \Delta l_{1.5} + \Delta
l_{1.5}\times \Delta M(a) \sim 0.89$ mas. The agreement with the plots
in the two bottom panels of Figure \ref{fig-astrom} is very good.  Figure
\ref{fig-astrom} also shows that the small difference (for an S2-like
orbit) between the purely relativistic perturbation and the combined
relativistic plus Galactic perturbation is smaller than astrometric
capabilities --- even in the case when the extended stellar cluster is
relatively massive.

\begin{figure}
\epsscale{0.8}
\plotone{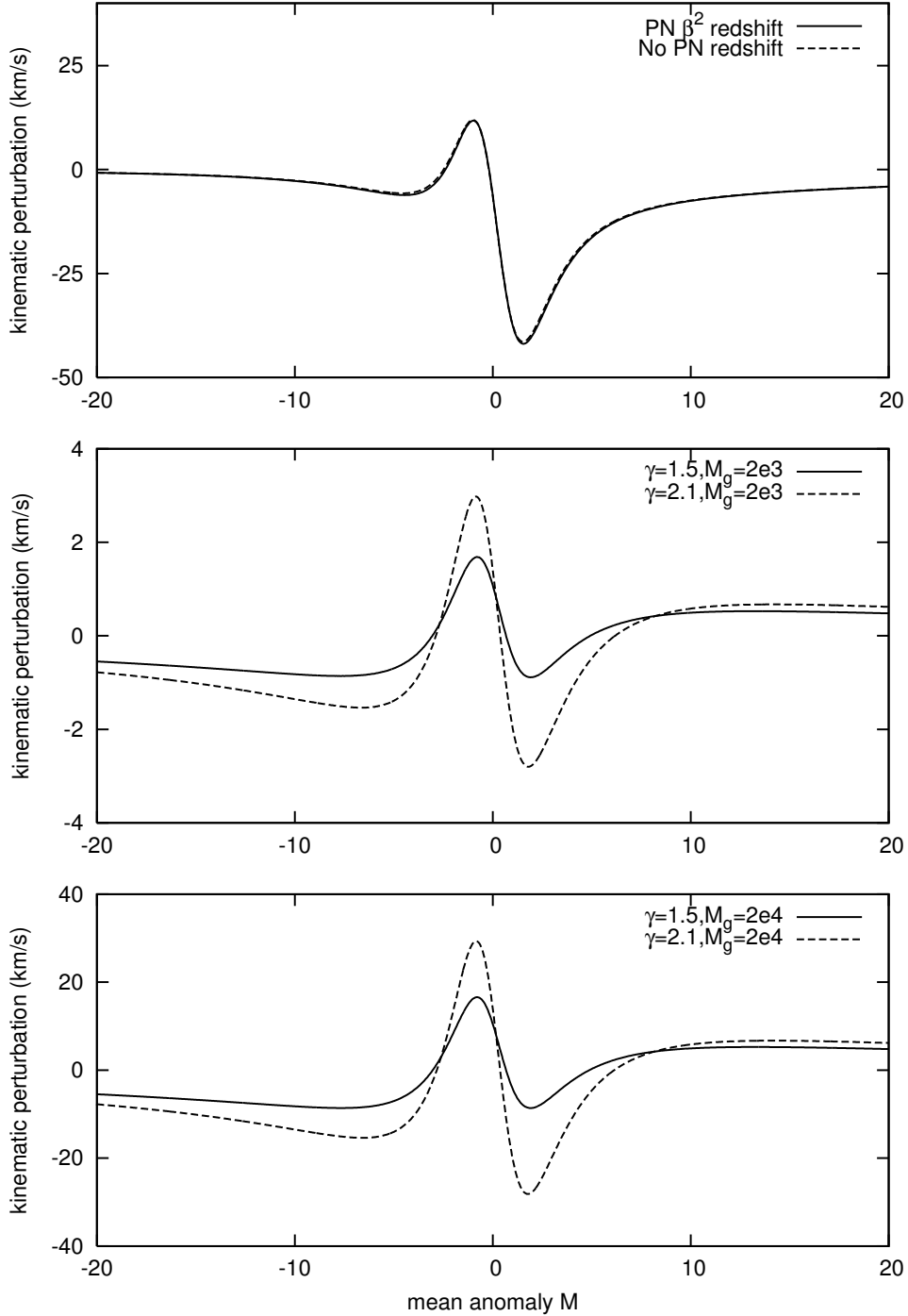}
\caption{Kinematic perturbations. As in
  Figure~\ref{fig-astrom}, the upper panel is from post-Newtonian
  effects only, the middle panel from the model Galactic perturbations
  only, and the lower panel combines both. Only the 10\% of the orbit
  around pericenter passage is shown here. The kinematic perturbation
  due to PN terms on a S2-like orbit appears measurable with current
  spectroscopic resolution, $\delta v \sim 10$ km/s.}
\label{fig-kinem}
\end{figure}

Figure~\ref{fig-kinem} shows the kinematic effect of relativistic and
Galactic perturbations.  As in the previous figures, the perturbations
are shown as a function of mean anomaly.  It can be seen that the
relativistic kinematic perturbation on a S2-like star is
strong enough to be detected by current instruments, for which $\delta
v \sim 10$ km/s. This is not the stronger case as the S14 star has a
closer pericenter passage and thus suffers a stronger (by a factor of
$\sim 5$) kinematic perturbation.

In summary, three more things are evident from these figures that are worth 
commenting on.
\begin{itemize}
\item For most of the known S-stars, the Galactic perturbations could be
  comparable to the relativistic effects, and even rather similar in
  time dependence. Perturbations on S2 and S14 are, however, essentially 
  dominated by relativistic effects. (Relativistic perturbations could
  dominate 
  even more in stars further in, if such stars are discovered; this, combined
  with shorter periods, would make detecting the signature of the relativistic
  effects much neater.)  It may be hoped, however, that even for the
  already-discovered stars, that the known and very specific form of the
  relativistic effects could
  enable them to be extracted, but it remains to be demonstrated.
\item Kinematics perturbations are concentrated near pericenter
  passage, whereas astrometric perturbations are of the same order
  throughout the orbit.  For highly eccentric orbits, the the
  astrometric perturbation $\Delta l$ has two different origins: near
  pericenter, $\Delta l$ is due to $\Delta v$, which shifts the phase
  of the orbit; near apocenter, $\Delta l$ comes precession
  $\Delta\omega$, amplified by the lever arm of the orbit.
\item Although the speed is maximal at pericenter, the kinematic
  perturbation is maximal (and can be much greater) at a phase before
  or after.
\end{itemize}

\section{Comparison with other post-Newtonian formulations}

In computing the preceding numerical results, we have made some
non-trivial choices of convention.
\begin{enumerate}
\item We have worked in what may be called be the Boyer-Lindquist
  gauge, which for zero spin reduces to the standard gauge of the
  Schwarzschild spacetime.
\item Our independent variable is the affine parameter, which is
  proportional to the proper time.
\item Our kinematic variables are the canonical momenta, and not
  derivatives of the coordinates.  In particular, when computing
  Keplerian elements, we have fed momentum and not velocity values
  into the classical formulas.
\end{enumerate}
Different choices can lead to some surprising differences in the
results.  Of course, observable quantities must not change.  But for
abstract quantities such as the instantaneous $a$, even the sign of
the post-Newtonian effect can switch.  We now explain how to convert
to other conventions, considering for simplicity only the leading
order Schwarzschild effects.

For a weak-field Schwarzschild spacetime, the most common gauge choice
is the harmonic form, where the metric is
\begin{equation}
ds^2 = - \left( 1 - \frac2r + \frac2{r^2} \right) dt^2
       + \left( 1 + \frac2r \right) d\Fx^2 .
\end{equation}
Working out $\frac12 g^{\mu\nu}$ we readily derive the Hamiltonian
\begin{equation}
H = - \frac12 \left( 1 + \frac2r \epsilon^2 + \frac2{r^2} \epsilon^4 \right)
      \pt^2 + \frac12 \left( \epsilon^2 - \frac2r \epsilon^4 \right) \Fp^2
    + O(\epsilon^6) .
\label{Hharmonic}
\end{equation}
with $\epsilon$ labeling orders as before, through the replacements
\begin{equation}
r \rightarrow \epsilon^{-2} r \qquad \Fp \rightarrow \epsilon \, \Fp .
\end{equation}
Since there is no explicit dependence on $t$, the conjugate momentum
$\pt$ will be constant.  As before, the initial value of $\pt$ just
sets the units of the affine parameter $\tau$, but if we set
\begin{equation}
\pt = 1 + \left( \frac{\Fp^2}2 - \frac1r \right) \epsilon^2 
    + \left( \frac1{2r^2} - \frac{3\Fp^2}{2r} - \frac{\Fp^4}8 \right)
      \epsilon^4
\label{H3D}
\end{equation}
initially then $H=-\frac12+ O(\epsilon^6),$ and the affine parameter
will equal the proper time along the orbit, to the given order.

As an aside, in Hamiltonian dynamics there is another possible
interpretation of (\ref{H3D}): we can take the function $\pt(\Fx,\Fp)$
as a Hamiltonian in its own right, having three degrees of freedom and
$t$ as the independent variable.  \cite{1994AJ....108.1962S} used this
Hamiltonian to incorporate the leading-order post-Newtonian effects
into a symplectic algorithm for long-term integration of planetary
orbits.

Considering now the coordinate velocity $\Fv\equiv d\Fx/dt$ we have
\begin{equation}
\epsilon\,\Fv 
    = \left( \frac{dt}{d\tau} \right)^{-1} \left( \frac{d\Fx}{d\tau} \right)
    = \left( \frac{\partial H}{\partial\pt} \right)^{-1}
      \left( \frac{\partial H}{\partial(\epsilon\Fp)} \right)
\end{equation}
which for the Hamiltonian (\ref{Hharmonic}) gives
\begin{equation}
\Fv = \pt^{-1} \left( 1 + \frac2r \epsilon^2 \right)^{-1}
               \left( 1 - \frac2r \epsilon^2 \right) \Fp
    + O(\epsilon^4) .
\end{equation}
Substituting from (\ref{H3D}) and rearranging, we have
\begin{equation}
\Fv = \left( 1 - \left( \frac{\Fp^2}2 + \frac3r \right) \epsilon^2 \right)
      \Fp + O(\epsilon^4) .
\label{velmodulation}
\end{equation}
Thus while $\Fx\times\Fp$ is a constant of motion, in $\Fx\times\Fv$
the modulation (\ref{velmodulation}) appears \citep[cf.\ the leading
term in Equation 4.9 in][taking their $\nu=0$ in the test particle limit]{2003CQGra..20..755B}. 

As noted above, for the main results of this paper, we have computed
orbital elements using the ``momentum convention'', as
\begin{equation}
- \frac1{2a} = \frac{\Fp^2}2 - \frac1r \qquad
a(1-e^2) = |\Fx\times\Fp|^2
\end{equation}
and so on.  On the other hand, if the ``velocity convention''
\begin{equation}
- \frac1{2a} = \frac{\Fv^2}2 - \frac1r \qquad
a(1-e^2) = |\Fx\times\Fv|^2
\end{equation}
is adopted instead, it follows from (\ref{velmodulation}) that the
orbital elements can change at $O(\epsilon^2)$.

\begin{figure}
\epsscale{0.8}
\plotone{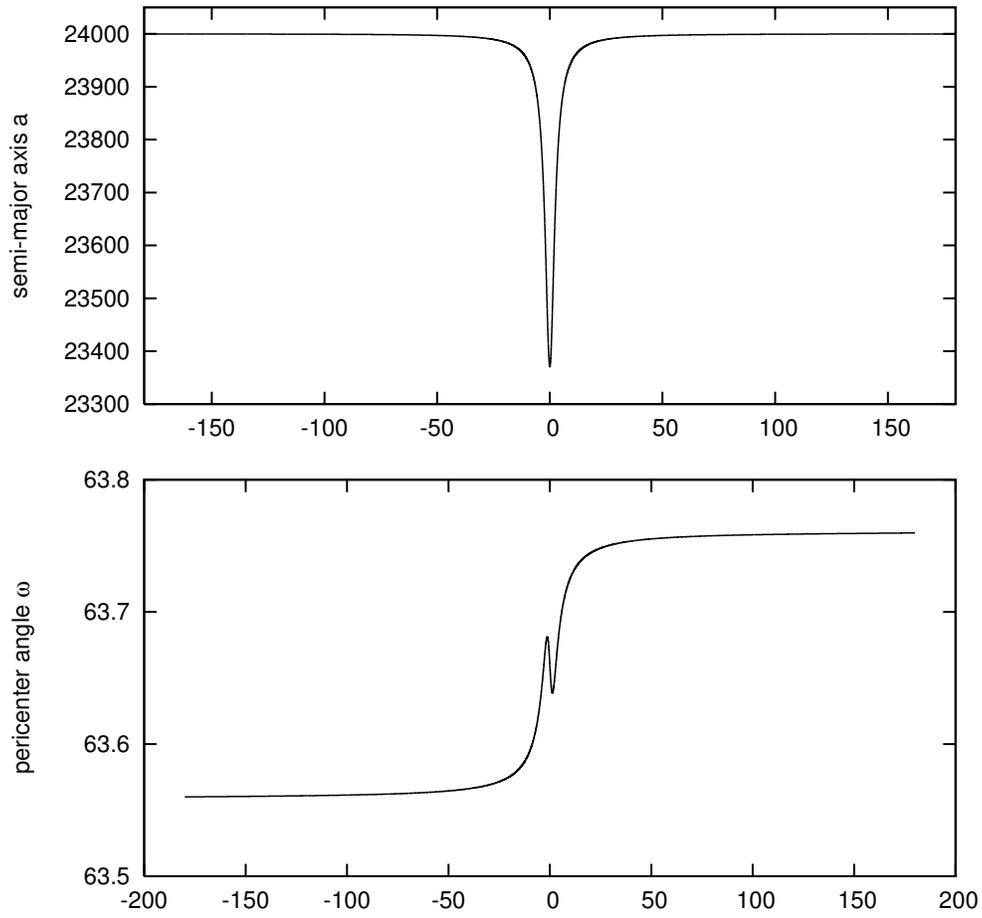}
\caption{Perturbation of the (osculating) orbital elements of an
  S2-like orbit due to post-Newtonian terms as in Figure~\ref{fig-elem}, 
  but now computing the orbital elements using the ``velocity convention''. 
  There are noticeable differences with Figure~\ref{fig-elem}: (i) the perturbation
  on the semi-major axis changes its sign; (ii) $\omega$  now shows a small oscillation 
  around  pericenter.
 }
\label{fig-elem-velconv}
\end{figure}

Figure~\ref{fig-elem-velconv} shows the perturbations of the orbital
elements  (as in the top two panels of Figure~\ref{fig-elem}), but now
computed by following the ``velocity convention''. It can be seen that
the semimajor axis perturbation changes its sign; while $\omega$ 
now shows a small oscillation around pericenter.

\section{Conclusions}

The prospect of detecting general relativistic effects in the S-stars
near the Galactic center has recently aroused interest.  To help gain
more insight into these small but exciting effects, we have derived a
simple formulation for the orbit equations and an algorithm for
numerically integrating them, in which the post-Newtonian dynamics
appears clearly as perturbations of the Kepler problem, and then
examined the size of the perturbations.

The post-Newtonian Hamiltonian is (\ref{Hpn}) or equivalently
(\ref{Hpn1}-\ref{Hpnf}).  These look like and are fairly simple
generalizations of the Keplerian Hamiltonian, the main difference
being that time is a coordinate and the affine parameter is an
independent variable.  For numerical integration we adapt the
variable-timestep symplectic integrators recently developed for
cometary or other highly eccentric orbits in the solar system.

For photons a somewhat different approximation applies than for
(comparatively) slow-moving stars, and we derive the Hamiltonian
(\ref{Hpn-null}).  It leads to gravitational redshift and related
effects, but we leave the computation of these for future work.

With the orbit integrator we proceed to compute the effects of the
post-Newtonian terms on the orbital elements, sky position, and
kinematics, taking the orbit of S2 as an example. For S2, general 
relativity implies an astrometric effect of
tenths of a mas and tens of km/sec in kinematics.  Two surprising
features are: (i)~the astrometric effect is of the same order near
pericenter as at apocenter, and (ii)~the kinematic effect is greatest
near (but not at) pericenter.

Newtonian perturbations due to other masses in the Galactic-center
region are unkown, but could plausibly be of the same order for the
stars so far know.  Disentangling the relativistic contribution from
the Galactic perturbations may be the hardest problem in practice.  It
would be necessary to fit simultaneously for post-Newtonian effects
and Galactic perturbations, without knowing the specific form of the
Galactic perturbation in advance.  Whether this is achievable is an
open question.  Simulations of the modeling pipeline are needed to get
a clear answer, but the known and very specific time-dependence of the
post-Newtonian effects suggest that we can be optimistic.

\acknowledgements We thank the referee, Clifford Will, for confronting 
some of our numerical results with linear perturbation theory, leading 
eventually to the comparison with standard post-Newtonian theory that
forms Section 5 of the paper.

MP acknowledges support by DLR (Deutsches Zentrum f\"ur Luft- und Raumfahrt).

\bibliographystyle{apj}
\bibliography{ms.bib}

\end{document}